\documentstyle[a4p,epsfig,cite,12pt]{article}
\begin{document}

\def\z{${\rm Z}^0$}
\def\ep{e$^+$e$^-$}
\def\ezh{{{\ep} $\to$ {\z} $\to {hadrons}$}}
\def\ezhc{{{\ep}$\to${\z} $\to{hadrons}$}}
\def\eh{{{\ep} $\to {hadrons}$}}
\def\ie{{\sl i.e.}}
\def\eg{{\sl e.g.}}
\def\ea{{\sl et al.}}
\def\vrs{{\sl vs.}}
\def\col{Collab.}
\def\JT{{\sc Jetset}}
\def\OP{OPAL}
\def\DE{DELPHI}

\def\al{\langle}
\def\ar{\rangle}
\def\ka{\kappa}

\def\vs{\vspace*}
\def\hs{\hspace*}
\def\bea{\begin{eqnarray}}
\def\eea{\end{eqnarray}}
\def\be{\begin{equation}}
\def\ee{\end{equation}}
\def\la{\label}
\def\bga{\left( \begin{array}}
\def\ena{\end{array} \right)}
\def\nwp{\newpage}
\def\ct{\cite}
\def\bi{\bibitem}
\def\ni{\noindent}
\def\fn{\footnote}

\def\pT{p_T}
\def\phi{\Phi}
\def\yf{y$$\times$$\phi}
\def\yf{y$$\times$$\phi}
\def\yp{y$$\times$$\pT}
\def\fp{\phi$$\times$$\pT}
\def\3d{y$$\times$$\phi$$\times$$\pT}
\def\lsim{\:{\stackrel{<}{_\sim}}\:}

\date{}

\def\jour#1#2#3#4{{#1} {#2} (19#3) #4}
\def\jourpt#1#2#3{{#1} (19#2) #3}
\def\jourm#1#2#3#4{{#1} {#2} (20#3) #4}
\def\PRp{Phys. Reports}
\def\PRD{Phys. Rev. {D}}
\def\PRC{Phys. Rev. {C}}
\def\PRL{Phys. Rev. Lett. }
\def\AP{Acta Phys. Pol. {B}}
\def\ZP{Z. Phys.  {C}}
\def\EPJ{Eur. Phys. J. {C}}
\def\IJ{Int. J. Mod. Phys. {A}}
\def\NIM{Nucl. Instr. Meth. {A}}
\def\CP{Comp. Phys. Comm.}
\def\TPS{Theor. Phys. Suppl.}
\def\PL{Phys. Lett.  {B}}
\def\NPA{Nucl. Phys.  {A}}
\def\NPB{Nucl. Phys.  {B}}
\def\MPL{Mod. Phys. Lett. {A}}
\def\JP{J. Phys. {G}}
\def\NC{Nuovo Cim.}
\def\UFN{Physics-Uspekhi}
\def\HEPC{High En. Phys. \& Nucl. Phys.}


\renewcommand{\Huge}{\huge}
\textheight=24.5cm
\pagestyle{empty}
\hs{4.cm}

\vs{1cm}
\begin{flushright}
TAUP $2649-2000$ \\
\vspace{2mm}

\today
\end{flushright}
\begin{center}
\vspace{9mm}
{\Large \bf 
Mass and Transverse Mass Effects\\ 
on \\ 
\vspace{2mm} 

the Hadron Emitter Size \\  
}
\vspace{5mm}

\bigskip
\bigskip
{\large Gideon Alexander\fn{Email address: alex@lep1.tau.ac.il}}
\vspace{4mm}

{\small \it School of Physics and Astronomy,\\
The Raymond and Beverly Sackler Faculty of Exact Sciences,\\
Tel-Aviv University, IL-69978 Tel-Aviv, Israel}
\vspace{1mm}

\vs{2.8cm}
\end{center} 

\medskip


\begin{abstract}
\vs{.7cm}
\ni
We investigate the dependence of the longitudinal
emitter dimension $r_{||}$ of identical bosons, produced in the 
hadronic Z$^0$ decays, on their transverse mass $m_T$ obtained
from 2-dimensional Bose-Einstein correlations (BEC)
analyses. We show that
this dependence is well described by the expression
$r_{||}=c\sqrt{\hbar\Delta t}/\sqrt{m_T}$, 
deduced from the uncertainty relations, setting $\Delta t$ to be
a constant 
of the order of $10^{-24}$ sec. This equation is
essentially identical to the one previously applied to the 
1-dimensional BEC results for the emitter
radius dependence on the boson mass itself. It is further shown 
that a very similar
behaviour exists also for the dependence of the
interatomic separation in  
Bose condensates on their atomic masses when they are at the same
very low temperature.  

\end{abstract}
\vspace{6mm}

\begin{center}
{\it (Submitted for publication)}
\end{center}

\vs{2.cm}
{
\ni

\ni
}
\nwp
\thispagestyle{empty}


\pagestyle{plain}
\setcounter{page}{1}

\section{Introduction}
Bose-Einstein Correlations (BEC) of identical bosons, produced in
multihadron final states of high energy particle interactions,
have been analysed for some 40 years \cite{review}. 
Many BEC analyses utilised  
pairs
of identical charged pions produced in multihadron final
states where the emitter has often been assumed to be 
a sphere with a Gaussian distribution. The experimental results
have been then subjected to a Coulomb correction to account for
the repulsive force between the equally charged particles. The kinematic
variable frequently used, and still in use today, is defined by\\
$$Q=\sqrt{-(q_1 - q_2)^2}\ ,$$
where $q_1$ and $q_2$ are the four momenta of the two identical
hadrons. In the limit of $Q\ \rightarrow\ 0$ 
the two identical bosons
are occupying the same lowest energy ground state defined 
in their centre of mass system.
To observe the effect of the BEC the experimental 
$Q$ distribution is divided by  
a corresponding reference
sample distribution which is selected so as to be, as much as possible,  
identical to the data sample in all its features 
but void of Bose-Einstein statistics effects. 
The distribution of this ratio
is in general
described by the expression
\begin{equation}
C_2(Q)=1+\lambda e^{-Q^2r^2}\ ,
\label{ctwo}
\end{equation}
where $r$ measures the average distance between the two boson when
they are predominantly in an s-wave. This $r$ value, extracted from
$C_2(Q)$ as $Q$ approaches zero,   
is taken to represent the dimension of the hadron emitter. The 
factor $\lambda$ in Eq. \ref{ctwo}, which can vary between 
$0 \leq \lambda \leq 1$, 
is the strength of the effect which depends on 
the chaoticity of the 
emitter and on the purity of the measured data sample.
In $e^+ e^-$ annihilations $r$ was found to be 
within the range of 0.7 to 1.0 fm (see e.g. Ref. \cite{report}), 
essentially independent of the centre of mass energy $\sqrt{s_{ee}}$.\\ 

\noindent
In recent years the BEC analyses in  $e^+ e^-$ annihilations have 
been extended in
several directions. Among them, the search for the so called
higher order BEC namely, of three or more genuine identical hadrons
correlations; the search for 
deviation from
an ideal spherical emitter and to studies aimed to
determine whether
the correlation dimension is a function of the
hadron mass.\\
 
\noindent
In Ref. \cite{rm} it was first pointed out that
in  $e^+ e^- \to Z^0 \to hadrons$ the measured dimension $r$ values 
are a 
decreasing function of
the hadron mass $m$ (see Fig. 1). This observation was first deduced 
from the measured $r$ values obtained from BEC analyses 
of identical $\pi^{\pm} \pi^{\pm}$ and 
$K^{\pm} K^{\pm}$ boson-pairs. That indeed  
$dr(m)/dm < 0$ was significantly strengthened   
by the recent emitter size 
measurements \cite{rlambda} of the $\Lambda \Lambda$ and
$\bar \Lambda \bar \Lambda$ pairs.
These last measurements utilised a method proposed in Ref. \cite{lipkin}
where the mixture of S=0 and S=1 spin states can be determined for
the hyperon-pair as a function of their centre of
mass (CM) energy. The onset of the Pauli exclusion principle, as the
CM kinetic energy decreases to zero and the s-wave of the systems
dominates, determines the $r(m_{\Lambda})$ value 
which was found to be of the order
of 0.15 fm.\\   

\begin{figure}[h]
\centering{\epsfig{file=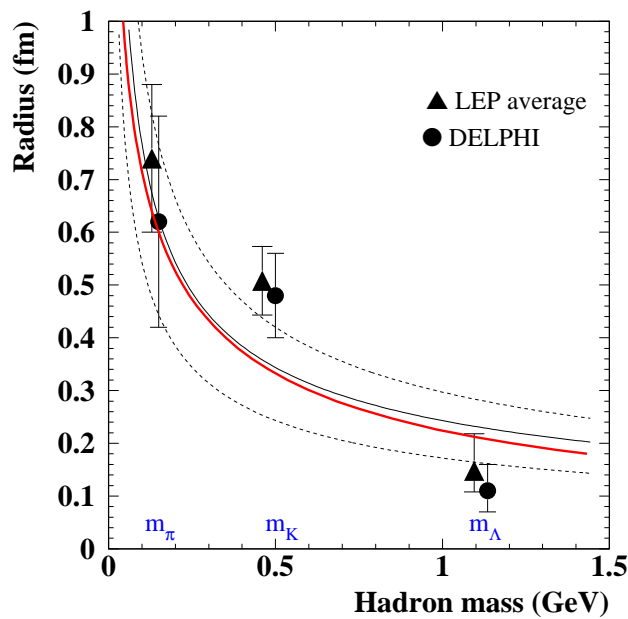,height=9cm}}
\caption{The measured emitter radius $r(m)$  
as a function of the hadron mass determined from BEC analyses 
using hadronic Z$^0$ decay events at LEP1
(taken from Ref. \cite{report}). The continuous thin line is the prediction
from the Heisenberg uncertainty relations setting $\Delta t = 10^{-24}$ sec,
the upper and lower dashed lines correspond respectively to the  
$\Delta t$ values of $1.5\times 10^{-24}$ sec and $0.5\times 10^{-24}$ sec.
The continuous thick line is derived from the virial theorem 
assuming Local Parton Hadron Duality and using a general QCD potential.  
}
\label{fig_rm}
\end{figure}

\noindent
Whereas the experimental findings that $r(m_{\pi})$ is somewhat
larger than $r(m_K)$, but still equal within errors, may still
be consistent with the 
string fragmentation model, 
although in its basic form it expects $r(m)$ 
to increase with $m$  \cite{lund}, 
the much smaller value obtained for $r(m_{\Lambda})$
poses a challenge to the model \cite{moriond}. 
At the same time however, 
it was shown in Ref. \cite{rm} 
that by applying the Heisenberg uncertainty relations, 
one can
derive an expression for $r(m)$ which decreases as $m$ increases, namely:
\begin{equation}
r(m)\ =\ \frac{c\sqrt{\hbar \Delta t}}{\sqrt{m}}\ . 
\label{uncertainty}
\end{equation}
Taking for  $\Delta t$ the value
$10^{-24}$ sec to represent the time scale of 
the strong interactions sector, 
independent of the hadron mass, one obtains 
the continuous thin line in Fig. \ref{fig_rm} which follows rather well
the trend of the $r$ values measured in the LEP1 data. 
A fit of Eq. \ref{uncertainty} to the data yields for $\Delta t$ 
the value $(1.2 \pm 0.3)\times 10^{-24}$ sec. The 
continuous thick line in Fig. \ref{fig_rm}, which is
almost identical to the one obtained from the uncertainty relations,
was derived from
the virial theorem assuming Local Parton Hadron
Duality \cite{lphd} using a general QCD potential
\cite{qcdp}.\\ 

\noindent
The effective range   
of the two-pion sources was also estimated in 2-dimensional BEC analyses, 
in heavy-ion collisions \cite{murray} and in the hadronic Z$^0$ decays
\cite{transverse},
as a function of the transverse mass $m_T$ of the pion-pair,
defined as
\begin{equation}
m_T\ =\ 0.5\times  
\left (\sqrt{m^2\ +\ p_{1,T}^2}\ +\ \sqrt{m^2\ +\ p_{2,T}^2}\ \right
)\ .
\label{eqmt}
\end{equation}
\noindent
Here $p_{1,T}$ and $p_{2,T}$
are the transverse momentum of the two identical bosons  
in the longitudinal centre of mass system (LCMS) \cite{lcms}.  
As can be seen e.g. in Fig. \ref{fig_rl}, the results of these studies
show also a decrease of the longitudinal range 
$r_z\ (\equiv r_{||})$ of the $\pi\pi$ system
as $m_T$ increases. 
Moreover, the
behaviour of $r_z^{\pi}(m_T)$ follows very closely the dependence
of $r(m)$ 
which is a function of the hadron mass itself.\\

\noindent
These findings pose the obvious question 
why the values of $r(m)$
and $r_z^{\pi}(m_T)$ essentially coincide when 
$m\ =\ m_T(\pi)$.
Also of interest is the question whether both or
one of the two quantities, $m$ and $m_T$, are the basic variables  
on which the hadron emitter dimension depends on and thus should play
an integral part in any model describing multi-hadron production.\\

\noindent
In the quest to understand the interrelation between $r(m)$ 
and $r_z^{\pi}(m_T)$ 
we explore in Section 2 the possibility that the
dependence of $r_z$ on $m_T$ can also be described in terms
of the Heisenberg uncertainty relations. Next in Section 3  
we turn to   
another phenomenon related to
the Bose statistics namely the Bose-Einstein
Condensation. Here we show that at a fixed very low
temperature the dependence of
the interatomic atomic separation on the mass of the
condensates atoms is proportional to $1/\sqrt{m_{atom}}$. 
Finally a summary 
is presented in Section 4.

\section{The longitudinal dimension dependence on {\LARGE\em m}$_T$}
Most of the BEC analyses were carried out 
under the assumption that the emitter
size is a Gaussian sphere. The possibility that the space-time
extend of the particle emission region deviates from a sphere and in
fact is characterised by more than one dimension has been recently proposed 
\cite{theor}. 
In particular
the Lund group developed for the BEC a model based on a quantum
mechanical interpretation of the string fragmentation probability
\cite{beclund}. In this model the correlation length 
in the longitudinal string direction should be larger than the
corresponding range in the transverse direction.\\ 

\noindent
The experimental analyses, most of which were carried out
with identical charged pions, have utilised the  
longitudinal centre of mass system. This coordinate
system is defined for each pair of
identical pions as the system in which the sum of the pion-pair
momenta $\vec{p}_1 + \vec{p}_2$, referred to as the 'out' axis, is 
perpendicular
to the 'thrust' (or jet) direction defined as the z-axis. 
The momentum difference of the pion-pair $\vec{Q}$ is then resolved into 
the longitudinal direction
$Q_z\ \equiv \ Q_{||}$ parallel to the thrust axis, $Q_{out}$ is
collinear with the pair momentum sum and the third axis $Q_{side}$,
is perpendicular to $Q_z$ and $Q_{out}$. In this system the projections
of the total momentum of the particle-pair onto the longitudinal and
side directions are equal to zero. In particular $p_{1,z} = -p_{2,z}$,
where the index 1 and 2 refer to the first and second pion, so that 
$Q_z\ =\ p_{1,z} - p_{2,z}\ =\ 2p_{1,z}\ =\ 2p_z$. The difference in the
emission time of the pions couples to the energy difference between
the particles only in the $Q_{out}$ direction.  
%
In a 2-dimensional analysis one defines the transverse component of Q
by the relation
$$Q^2_T\ =\ Q^2_{out}\ +\  Q^2_{side}\ .$$
Thus the correlation function, which is fitted to the data, is of the
form
\begin{equation}  
C_2(Q_z, Q_T)\ =\  1\ +\ \lambda e^{-(r^2_zQ^2_z\ 
+\ r^2_TQ^2_T)}\ , 
\label{twodim}
\end{equation}
where $r_z$, estimated from Eq. \ref{twodim}
as $Q_z$ approaches zero,  
is the longitudinal geometrical radius and $r_T$
is a mixture of the transverse radius and the emission time.
The experimental findings, both in heavy ion collisions \cite{na}
and in $e^+ e^-$ annihilations \cite{rlrt}, verified the theoretical 
expectations that $r_T/r_z$ is significantly smaller than one.\\ 
\begin{figure}[h]
\centering{\epsfig{file=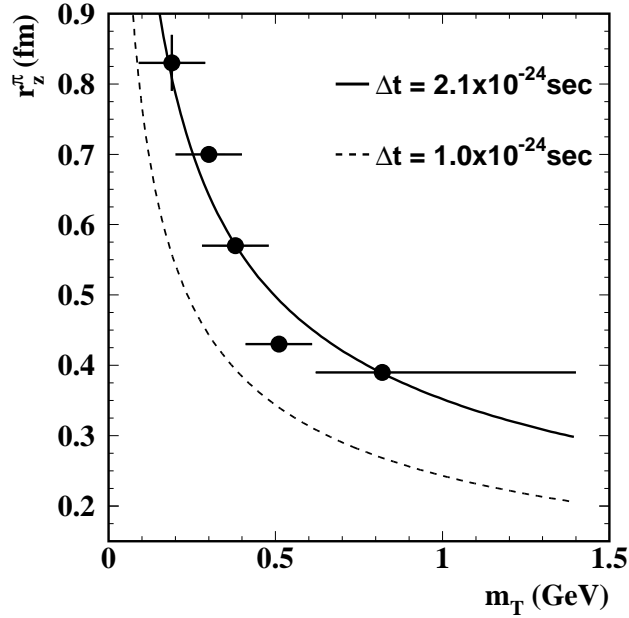,height=9cm}}
\caption{Preliminary results of DELPHI \cite{preli} 
for the dependence of the longitudinal emitter dimension $r_z^{\pi}$ 
on the transverse mass  $m_T$ 
in hadronic Z$^0$ decays.
The data is compared with the expression for $r_z$ given in
Eq. \ref{finalmt} setting $\Delta t$ to the best fitted value of 
2.1 $\times 10^{-24}$ sec (continuous line) and the expectation for 
1.0  $\times 10^{-24}$ sec (dashed line).}  
\label{fig_rl}
\end{figure}

\noindent
In the frame work of the azimuthally symmetric sources of 
pair of identical bosons \cite{heinz_bec} another variable
of the emission function is considered namely,
the transverse mass $m_T$ defined by Eq. \ref{eqmt}. 
Preliminary results of DELPHI \cite{preli} 
concerning the dependence of $r_z^{\pi}(m_T)$ on $m_T$, 
measured for identical charged pion pairs present in the hadronic Z$^0$
decays, is shown in Fig. \ref{fig_rl}. As can be seen,  $r_z^{\pi}(m_T)$ 
decreases with $m_T$ in a very similar way to the decrease of
$r(m)$ as the mass $m$ increases. In fact the continuous line in the
figure, which is drawn according to 
Eq. \ref{uncertainty} replacing $r(m)$ by $r_z^{\pi}(m_T)$ and $m$ by $m_T$,
describes well the $r_z$ measurements using for $\Delta t$ 
the value $2.1 \times 10^{-24}$ sec. 
That this is the case is not surprising once one realises that 
$\Delta r_z$, the longitudinal distance and $\Delta p_z$, 
the difference in the longitudinal momentum
of the two hadrons in the LCMS, are conjugate observables which obey
the uncertainty principle
\begin{equation}
\Delta p_z \Delta r_z\ =\ \hbar c\ ,
\label{drdp}
\end{equation}
Here $\Delta p_z$ is
measured in GeV, $\Delta r_z\ \equiv\ r_z$ is given in fermi units and 
$\hbar c\ =\ 0.197$ GeV fm. In the LCMS one has 
$$\Delta p_z r_z\ =\ 2\mu v_z r_z\ =\ p_z r_z\ =\ \hbar c$$  
where $\mu\ =\ m/2$ is the reduced mass of the two identical hadrons of mass
$m$ and the longitudinal velocity $v_z$ of these hadrons. 
Thus
\begin{equation}
r_z\ =\ \frac{\hbar c}{p_z}\ . 
\label{rz}
\end{equation}

\noindent
Simultaneously we also utilise the
uncertainty relation expressed in terms of energy and time
\begin{equation}
\Delta E \Delta t\ =\ \hbar\ ,
\label{dedt}
\end{equation}
where the energy is given in GeV and $\Delta t$ in seconds. 
In as much that the total energy $E$ of the two-hadron system
is determined essentially only by their mass and their kinetic energy,
i.e. the potential energy can be neglected, and since in the LCMS\
$|p_{1,z}| = |p_{2,z}|$, one has
\begin{equation}
E\ =\ \sum_{i=1}^2 \sqrt{m^2 + p^2_{i,x} +  p^2_{i,y} +  p^2_{i,z}}\ =\ 
\sum_{i=1}^2 \sqrt{m^2_{i,T} +  p_z^2}\ ,
\label{deltae}
\end{equation}
where $m_{1,T}$ and $m_{2,T}$ are the transverse mass of the first and
second hadron.
As $Q_z$ decreases the longitudinal momentum $p_z$ vanishes 
so that we can, once  $0 \leq p_z^2 <  m^2_{i,T}$,
expand the hadron energy $E$ in terms of $p_z^2/m^2_{i,T}$ and 
retain only the two first terms,
\begin{equation}
E\ =\ \sum_{i=1}^2 m_{i,T} \sqrt{1\ +\ \frac{p_z^2}{m^2_{i,T}}}\ 
\approx\ \sum_{i=1}^2 m_{i,T} + \sum_{i=1}^2 \frac{p_z^2}{2 m_{i,T}}\ .
\label{expand}
\end{equation}
Next we order the identical bosons so that  $m_{1,T} \geq  m_{2,T}$ and define
$$ \delta m_T\ =\ \frac{m_{1,T} - m_{2,T}}{2}\ \geq 0\ ,$$
while $$m_T\ =\  \frac{m_{1,T} + m_{2,T}}{2}\ .$$ 
Inserting these relations into Eq. \ref{expand} one gets, after few
algebraic steps, that
\begin{equation}
E\ =\ 2 m_T\ +\ \frac{m_T p^2_z}{m^2_T - (\delta m_T)^2}\ . 
\end{equation}
As $m^2_T$ is larger than  $(\delta m_T)^2$, 
we finally get  
\begin{equation}
E\ \approx\ 2 m_T\ +\ \frac{p^2_z}{m_T}\ . 
\label{finale}
\end{equation}

\noindent
Since 2$m_T$ is not a function of $p_z$ it may be considered 
to stay fixed as $Q_z \to 0$, so that one has  
\begin{equation}
\Delta E \Delta t\ =\ \frac{p_z^2}{m_T}\Delta t\ =\ \hbar\ . 
\label{deltaf}
\end{equation}
Combining Eqs. \ref{rz} and \ref{deltaf}
one obtains
\begin{equation}
r_z(m_T)\ \approx\ \frac{c\sqrt{\hbar \Delta t}}{\sqrt{m_T}}\ .  
\label{finalmt} 
\end{equation}
This last equation is identical to the one derived in \cite{rm}
for the dependence of emitter dimension on the boson mass  
when $r(m)$ and $m$ are replaced by $r_z(m_T)$ and $m_T$. A fit of
Eq. \ref{finalmt} to the data shown in Fig. \ref{fig_rl} 
yields $\Delta t\ =\ (2.1 \pm 0.4)\times 10^{-24}$ sec
so that $r_z^{\pi}(m_T)\ =\  0.354/\sqrt{m_T(GeV)}$ fm.
This value is compatible with the value of 
$\Delta t\ =\ (1.2 \pm 0.3)\times 10^{-24}$ sec obtained in
\cite{rm} for $r(m)$ when one also takes into account the relatively wide  
spread of the 1-dimensional $\pi\pi$ BEC analyses results
for $r(m)$ obtained by
the four LEP1 experiments (see e.g. Ref. \cite{rm}). 
In heavy-ion collisions of S + Pb, at an energy of 200 GeV per nucleon,  
the longitudinal range $r_z^{\pi}(m_T)$ 
was also observed
to be inversely proportional to the square root of $m_T$ 
\cite{heinz_bec} namely, 
$r_z^{\pi}(m_T)\ \approx\ 2/\sqrt{m_T(GeV)}$ fm. The
ratio between the proportionality factor of 2.0 and 0.354 may
well be accounted for by the difference in the extend of
the heavy ion target as compared to that of the $e^+ e^-$  
annihilation leading to hadronic Z$^0$ decays.\\    

\noindent
The dependence of the transverse dimension, $r_T^{\pi}$, on the 
two-pion transverse
mass $m_T$ has also been measured in the hadronic Z$^0$ decays
\cite{tr}. Here again the transverse range was found to decrease
as $m_T$ increases. However 
unlike $r_z$ which is a geometrical quantity, $r_T$ is a mixture of the 
transverse radius and the emission time so that an 
application of the uncertainty relations is not straightforward.
 
\section{Interatomic separation in Bose Condensates}

When bosonic atoms are cooled down, below a critical temperature $T_B$, 
the atomic wave-packets overlap and the 
equal identity of the particles becomes significant. 
At this temperature, these atoms undergo a 
quantum mechanical phase transition and form a Bose-Einstein (BE)
condensate, a coherent cloud of atoms all occupying 
the same quantum mechanical state.
This phenomenon, first predicted by
A. Einstein in 1924/5, is a consequence of 
quantum statistics \cite{einstein}.
Detailed theoretical aspects of the Bose-Einstein condensation
can be found in Ref. \cite{theory} and its  
up to date experimental situation is described in Ref. \cite{experiment}.
Concise summaries, aimed in particular to the non-expert, 
both of the experimental situation and the theoretical
background, can be found in Refs. \cite{ketterle,burnett}. 
To form Bose condensates one cools down, below the
critical temperature $T_B$,
extremely dilute gases so that the formation time of molecules and
clusters in three-body collisions is slowed down to seconds or
even minutes to prevent the creation of more familiar transitions into
liquid or even solid states.\\

\noindent
The existence of BE Condensation was first 
demonstrated in 1995 by three groups
\cite{discovery} in cooling down rubidium, sodium and lithium.
Typical temperatures where BE condensates occur are in the
range of 500 nK  to 2 $\mu$K with atom densities between
$10^{14}$ and $10^{15}$ cm$^{-3}$. The largest sodium condensate 
has about 20 million atoms whereas hydrogen 
condensate can reach even one billion atoms.\\

\noindent 
Let us consider a dilute homogeneous ideal gas of $N$ identical bosonic atoms
of spin zero,
confined in a volume $V$. These atoms occupy energy levels 
$\epsilon$, handled 
here as a
continuous variable, which are distributed according to the  
Bose-Einstein statistics. 
We further set the ground state to be
$\epsilon_0 = 0$. If $N_0$ is the number of atoms in this ground state
and $N_{ex}$ is the number of atoms in the excited states then
$N\ =\ N_0\ +\ N_{ex}$. For a 
homogeneous ideal gas of identical bosonic atoms it can be shown
\cite{book} that at a low temperature $T$ one has
\begin{equation}
N_{ex}=2.612\, V\left(\frac{2\pi mkT}{h^2}\right )^{3/2}\ ,
\label{exvn}
\end{equation}
where $V$ is the volume occupied by the atoms of mass $m$. 
Since $T_B$ is defined as
the temperature where almost all bosons are still in excited states,
we can, to a good approximation, equate $N$ with $N_{ex}$. That is
\begin{equation}
N = 2.612\, V \left(\frac{2\pi mkT_B}{h^2}\right )^{3/2}\ ,
\label{exnn}
\end{equation}
For $T< T_B$ one obtains from Eqs. \ref{exvn} and \ref{exnn}  
that the number of atoms $N_0$ which are in the condensate state is, 
\begin{equation}
N_0= N - N_{ex} = N\left[1- \left(\frac{T}{T_B}\right )^{3/2}\right ]\ ,
\label{nzeron}
\end{equation} 
where $T$ is the temperature of the atoms lying at the excited energy
states
above the condensate energy level $\epsilon_0 =0$. 
The atomic density of the Bose gas at very low temperatures,
$T/T_B << 1$ where $N \approx N_0$, is then given by
\begin{equation}
\rho=\
\frac{N}{V}=2.612\left(\frac{2\pi mkT}{h^2}\right )^{3/2}\ ,
\label{exdn}
\end{equation}
\noindent
where $k$ is the Boltzmann constant and
$\rho$, the atomic density, has the dimension of L$^{-3}$.
From this follows that $\rho^{-1/3}$ is the average interatomic 
separation in the Bose condensate. At the same time the thermal 
de Broglie wave length is equal to
\begin{equation}
\lambda_{dB}=\left(\frac{h^2}{2\pi mkT}\right)^{1/2}\ . 
\label{broglie}
\end{equation}
Combining Eqs. \ref{exdn} and \ref{broglie} one has
for the state of a Bose condensate the relation
\begin{equation}
\rho \lambda ^3_{dB} \approx 2.612\ .
\label{condition}
\end{equation}
Thus the average interatomic distance in a Bose condensate,
$d_{BE}$, is equal to
\begin{equation}
d_{BE} \equiv \rho^{-1/3} \approx \lambda_{dB}/1.378\ .  
\end{equation}

\noindent
Next we consider two different bosonic gases, having atoms with masses
$m_1$ and $m_2$, which are cooled down to the same very low
temperature $T_0$, below the critical temperature 
$T_B$ of each of them. In this case we will produce two Bose condensates 
with interatomic distances
\begin{equation}
d_{BE}(m_i)\ \approx 
\ \frac{\sqrt{2\pi}}{1.378}\left(\frac{\hbar^2}{m_ikT_0}\right)^{1/2}\
;\ \ \ i\ =\ 1,2\ .
\label{dbfinal}
\end{equation}  
From this follows that when two condensates are at the same
fixed temperature $T_0$ one has
\begin{equation}
\frac{d_{BE}(m_1)}{d_{BE}(m_2)} = \sqrt{\frac{m_2}{m_1}}\ ,
\label{mmm}
\end{equation}
which is also the expectation of Eq. \ref{uncertainty} 
for the dimension
dependence on the mass of the hadron produced in high energy 
reactions provided $\Delta t$ is fixed. Finally it is
interesting to note that in as much that one 
is justified to replace in Eq. \ref{dbfinal}, ~at a very low temperature, 
$kT_0$ by 
$\Delta E$ and use the uncertainty relation 
$\Delta E = \hbar /\Delta t$ one derives the expression for $r(m)$ 
as given by Eq. \ref{uncertainty} multiplied by the factor 
$\sqrt{2\pi}/1.378$.\\

\noindent 
In relating the condensates with the production 
of hadrons in high energy reactions one should however   
keep in mind that the interatomic separation proportionality
to $1/\sqrt{m}$ does not necessarily imply
that this should also be so for 
hadrons produced in high energy reactions. 
Common to both systems is their
bosonic nature which allow all hadrons (atoms) to occupy
the same lowest energy state. In addition the condensates are 
taken to be in
a thermal equilibrium state. Among the various models
proposed for the production of hadrons in high energy $e^+ e^-$
and Nucleon-Nucleon reactions
some attempts have also been made to explore the application of
a statistical thermal-like model \cite{becattini}. 
However, whether this approach will eventually prevail
is at present questionable. Finally the
condensates are taken to be in a  coherent state. In the case of the
of hadrons one is able to measure $r$ via BEC only if the chaoticity
factor $\lambda$ in Eq. \ref{ctwo} is different from zero 
i.e., only if the source
is not 100$\%$ coherent. However so far there is no evidence for
a dependence of $r$ on $\lambda$ apart from that introduced by
the correlated errors between $\Delta r$ and $\Delta \lambda$ 
produced by  
fitting Eq. \ref {ctwo} to the data.

\section{Summary}

It is shown that the longitudinal range of the two-pion emitter size
obtained from 2-dimensional BEC analyses of the hadronic Z$^0$ decays
as a function of the transverse mass,
is well described by the expectation
of the Heisenberg uncertainty relations using a constant value for
$\Delta t$ of the order of $10^{-24}$ seconds. As a consequence
it is not surprising that $r_z^{\pi}(m_T)$ has essentially the
same behaviour as the one observed experimentally for $r(m)$ which is
a function of the hadron mass itself. In both cases
the range, in fm, is equal to $(0.2-0.4)/\sqrt{m_T}$. This 
kind of dependence on the transverse mass is also seen in heavy-ion
collisions where it was found that
$r_z^{\pi} \approx 2/\sqrt{m_T}$.\\ 

\noindent
It is interesting to note that the interatomic separation of
atoms in different Bose condensates, having the same fixed 
temperature, is proportional to
$1/\sqrt{m}$, where here $m$ is the atom mass. This behaviour
is the same as found 
for the range between 
identical hadron-pair produced 
in high energy reactions when 
the time scale $\Delta t$ is fixed. This similarity can be traced back 
to
the close connection between the de Broglie wave length, applied to
the bosonic atoms in condensates, and the 
Heisenberg uncertainty relations, used here to connect in high energy 
reactions the range between identical hadrons to their mass.

\subsection*{Acknowledgements}
We would like to thank T. Cs\"{o}rg\H{o}, 
E. Levin, B. Reznik and E.K.G. Sarkisyan
for many helpful discussions. In particular our thanks are due to
I. Cohen for her continuous help in this work and her 
diligent proof reading of
the manuscript.
\label{sum}

\end{document}